\documentclass[fleqn,10pt]{wlscirep}

\title{Motif analysis in directed ordered networks and applications to food webs}

\author[1,2]{Pavel V.~Paulau}
\author[1]{Christoph Feenders}
\author[1,*]{Bernd Blasius}
\affil[1]{CvO University Oldenburg, ICBM, Carl-von-Ossietzky-Strasse 9-11, 26111 Oldenburg, Germany}
\affil[2]{Jade University of Applied Sciences, Ofener Strasse 16-19, 26121 Oldenburg, Germany}

\affil[*]{blasius@icbm.de}


\begin{abstract} 
The analysis of small recurrent substructures,
so called network motifs, has become a standard tool of complex network science to unveil the  design principles underlying the structure of empirical networks.
In many natural systems network nodes are associated with an intrinsic property according to which they can be ordered and compared against each other.
Here, we expand standard motif analysis to be able to capture the hierarchical structure in such ordered networks.
Our new approach is based on the identification of all ordered 3-node substructures and the visualization of their significance profile.
We present a technique to calculate the fine grained motif spectrum by resolving the individual members of isomorphism classes (sets of substructures formed by permuting node-order).
We apply this technique to computer generated ensembles of ordered networks and to empirical food web data, demonstrating the importance of considering node order for food-web analysis.
Our approach may not only be helpful to identify hierarchical patterns in empirical food webs and other natural networks, it may also provide the base for extending motif analysis to other types of multi-layered networks.
\end{abstract}
\begin{document}

\flushbottom
\maketitle
%
%
\thispagestyle{empty}   

%
%
%

\section*{Introduction}

The notion of complex networks has emerged in the last decades as an important framework for describing and investigating the organization of natural and living systems\cite{Newman2010}. 
A complex network can be regarded as any collection of units, or nodes, that are interacting as a system and that are connected by directed or undirected links. 
In this respect, complex networks serve as natural models to describe the structure of a diverse range of systems\cite{Newman2010,Blasius2011}, with examples from  social\cite{Borgatti2009}, economical\cite{Jackson2008}, technological\cite{Albert2004,Kaluza2010}, and biological systems\cite{Alon2003}.
The network perspective also plays an important role for describing the organization of ecological systems\cite{Jordan2004,Proulx2005}.
In particular, the network theoretic analysis of food webs has proven to be very useful to explore the properties of  `who eats whom' in ecosystems \cite{Dunne2009}.

Over the last years a wide variety of techniques has been put forward that allow to characterize the  structure and topology of complex networks\cite{Costa2007,Newman2010}.
These include statistical measures, such as centrality indices, that characterize the importance of single nodes or links in the network and network modules that capture the large-scale structure within the network.
These studies demonstrated that many real-world networks, and in particular biological networks, contain small network building blocks, the network motifs, that recur in a network more frequently than statistically expected\cite{Milo2002}.
The importance of network motifs relies largely on the observation that small subgraphs
indicate particular patterns of interactions between network nodes and thus may carry a regulatory or dynamic function.
The frequency distribution of small subgraphs in the network can by  visualized in the form of a significance profile or motif spectrum. 
This can be regarded as a fingerprint of the local network structure and allows
to systematically compare different networks and group them into superfamilies of networks with very similar significance profiles\cite{Milo2004}.
The analysis of motif spectra 
has recently gathered much attention as a useful concept to  
unveil universal design principles underlying the structure of complex networks, 
and it has been applied for the analysis of diverse networks\cite{Milo2002,Milo2004,Sporns2004,Itzkovitz2005,Kaluza2010}, 
including food webs\cite{Milo2002,Stouffer2007}.

Standard network characteristics, such as motif spectra, however 
cannot per se capture the structure of networks that have multiple
layers of complexity and therefore cannot be represented as a traditional graph.
Here, we propose an extension of standard motif analysis for a specific network type, which we denote as {\em directed ordered networks}.
Ordered networks are networks in which the nodes can be compared or related to each other by a binary ordering relation, $<$, that is independent of the graph topology.
That is, we assume that for every two nodes in the network it is known which node is `smaller' or `larger' than the other.
Ordered networks might reflect, for example, systems where the nodes are associated with an additional intrinsic property, such as size, fitness, importance, or geographic location\cite{Caldarelli2002}.
In these systems the nodes can be naturally ordered according to the value of their intrinsic state variable.
For example, individuals in a social network may be sorted according to their social status or financial income, (air)ports in a large-scale transportation network may be sorted according to their size or geographic location, and proteins in a regulatory network may be sorted according to their molecular mass or ubiquity among biota.
In fact, such intrinsic states arise naturally in many natural and technical systems and therefore most empirical networks can be regarded as being ordered.

Ordered relationships are also of fundamental importance for describing trophic species interactions in ecological food webs.
Here, species are sorted according to their body-size, accounting for the fact that it is more likely to find large predators feeding on small prey, than vice versa, i.e. large prey having small predators\cite{Williams2000,Williams2008,Williams2011}.
Empirical food webs have been constructed for diverse ecosystems \cite{Havens1992,Condit2000,Tuomisto2003,Thieltges2009,Lang2013} and statistical\cite{Williams2000,Williams2008,Williams2011,Dunne2009} as well as evolutionary\cite{Loeuille2005,Allhoff2013} 
models have been developed to understand mechanisms underlying them.
Food webs have been characterized by a variety of network theoretic measures, such as typical food chain length, clustering coefficient, or degree distribution\cite{Dunne2009}. 
In particular, motif analysis of empirical food webs has shown to be helpful to obtain a quantitative analysis of their local structure \cite{Milo2002,Stouffer2007,CamachoStoufferAmaral2007,Stouffer2010}.
However, non of these measures is able to incorporate the body-size related hierarchy 
pattern, i.e., the directed ordered structure, of food webs.

In this report, we develop a general framework for motif analysis in directed ordered networks.
We first propose the notion of directed ordered networks as 
directed networks where the nodes constitute a totally ordered set.
Next, we expand standard motif analysis to be able to capture the hierarchical structure in such systems.
Our new approach is based on the identification of all ordered 3-node substructures and the visualization of their significance profile.
Finally, we apply this technique to computer generated ensembles of ordered networks and to empirical food web data.
Thereby, we demonstrate that the extended motif analysis is a promising
technique to analyze hierarchical patterns and the role of body-size order in natural food webs and other complex networks.

\section*{Directed Ordered Networks}
\subsection*{Definition of Ordered Networks}

We define an {\em ordered network} as a graph, consisting of a collection of nodes that are connected by links, where the nodes constitute a totally ordered set.
This means that nodes can be linearly ordered, i.e., for any pair of non-identical nodes $i\neq j$ either the binary relation  $i<j$ or $j<i$ is true.
This binary relation is transitive, that is if $i<j$ and $j<k$ then also $i<k$.
When the nodes are ordered we denote the index of a node as its rank.
If all network edges have an orientation (i.e., the underlying network is directed) we speak of a {\em directed ordered network}.

Ordered networks arise naturally, but are not restricted to, the important situation where every node $i$ is associated with an intrinsic state $n_i \in \mathbb{R}$\cite{Caldarelli2002}.
In this case, the nodes can be embedded along a one dimensional niche axis and,
accordingly, we denote this intrinsic state as the node's {\em niche variable}.
Node sets with niche variables can be naturally ordered such that $n_i<n_j$ is equivalent to $i<j$. That is, nodes with smaller niche variables have smaller rank.
Depending on the specific context, nodes can be ordered according to any continuous niche variable that characterizes or is associated to the node, such as the size, fitness, importance, a dynamical state variable, a physical (e.g., temperature) or biological (e.g., population density) condition, or the geographic location. 
In this sense, ordered networks are ubiquitous in nature and, depending on the question, the same network may be ordered differently according to different niche variables.

\begin{figure}[ht]
\centering
\includegraphics[width=6.0cm,keepaspectratio=true,clip=true]{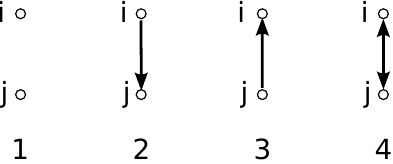} 
\caption{
Four possible relations between any two nodes $i$ and $j$ $(\mbox{with } i>j)$ in a directed ordered network: (1)~no~connection, (2)~downward connection, (3)~upward connection, (4)~bidirectional connection.
The node $j$ with smaller rank $j < i$ is plotted vertically below the node $i$ of higher rank.
} 
\label{PossibilitiesTwoOrderedDirected}
\end{figure}

The existence of a total ordering breaks the exchange symmetry between the network nodes and thus has important consequences for the network structure. 
In particular, in a directed ordered graph, any two nodes~$i$ and~$j$ are related in one of four possible ways (Figure~\ref{PossibilitiesTwoOrderedDirected}):
Two nodes may not be connected at all, there may be a unidirectional link downwards, a unidirectional link upwards (note that we always place nodes with smaller rank vertically below nodes with higher rank), or there may be a bidirectional link.
This distinction between downward connections (i.e., links pointing from nodes with a higher rank to a lower rank, Figure~\ref{PossibilitiesTwoOrderedDirected}, case $2$)
and upward connections (Figure~\ref{PossibilitiesTwoOrderedDirected}, case $3$) increases the combinatorial complexity of directed ordered networks compared to non-ordered networks and will be the base of our subsequent motif analysis.

\subsection*{Motif Analysis of Directed Ordered Networks}

Network motifs are small subgraphs, commonly corresponding to triplets of nodes, that are significantly overrepresented in a network\cite{Milo2002}.
To identify the motifs in a given network, the frequency of subgraphs in the network is compared to the expected number in an ensemble of randomized networks.
If the frequency of a given subgraph in the network is significantly larger than the mean frequency in a randomized ensemble, the subgraph is considered a network motif.

\begin{figure}[ht]
\centering
\includegraphics[width=14cm,keepaspectratio=true,clip=true]{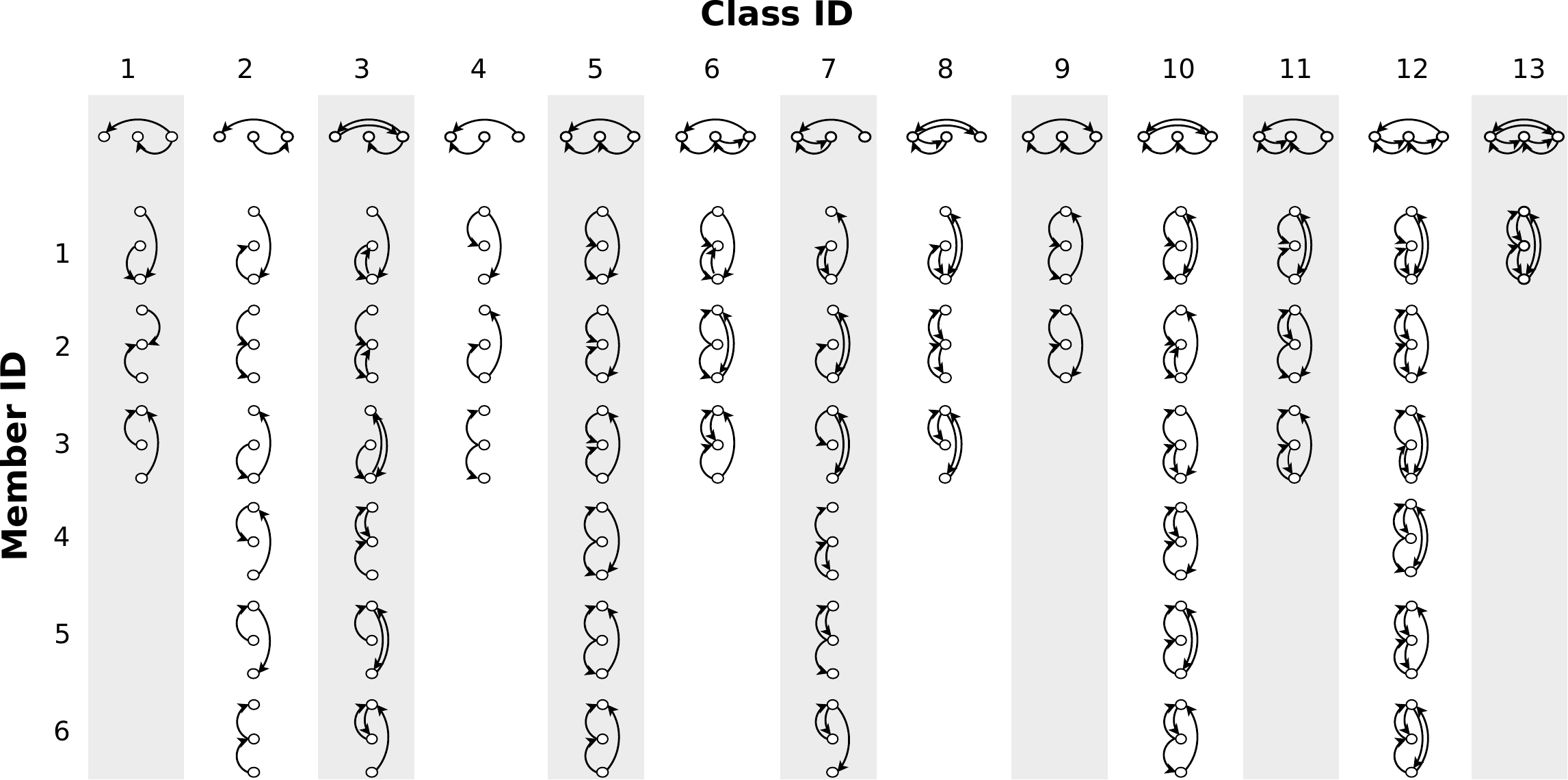}
\caption{
All unordered and ordered network motifs with 3~nodes.
The top row contains all 13 possible 3-node motifs when the node-order is not considered. 
Taking into account
node-order yields 54 ordered motifs (shown below the top row), which are arranged below their respective isomorphism class.
Thereby, each of the 13 unordered 3-node motif classes (labeled by their respective motif ID)
corresponds to 1 to 6 ordered  ordered 3-node motifs (labeled by their member ID).
For each ordered motif, the rank of nodes increases from bottom to top.} 
\label{figMotifsMap}
\end{figure}

In the following we develop an approach to extend the motif analysis to directed ordered networks.
Since in a directed ordered graph any two nodes~$i$ and~$j$ are related in one of four possible ways (Figure~\ref{PossibilitiesTwoOrderedDirected}), each triplet of nodes can be in one of $4^3=64$ possible configurations.
Ten of those configurations contain isolated nodes (i.e., nodes without any link) and will not be considered in the following.
The remaining $54$ connected ordered configurations, or substructures, are shown in Figure~\ref{figMotifsMap}.
When the hierarchy among nodes is neglected, the ordered substructures collapse to the well-known 13 classes from standard motif analysis\cite{Milo2002}.
Thereby, each of the 13 classes has a different size, i.e. it contains a different number of ordered 3-node substructures (Figure~\ref{figMotifsMap}).
For identification, we denote each substructure by a pair of indices~$(q,s)$, where $q=1\dots 13$ is the class ID and $1 \leq s \leq 6$ is the member ID.

In general, in directed ordered networks, downward links will have a different frequency than upward links.
This generalizes to substructures containing three nodes: their likelihood of appearance depends on the number of up- and downward connections.
We call the numbers of appearances, $\eta_{(q,s)}$, of all ordered substructures in a network {\em the spectrum of ordered 3-node substructures} or {\em motif spectrum} for short.
Unordered motif spectra can be retrieved by summing over member IDs
\begin{equation}
\eta_q=\sum \limits_{s} \eta_{(q,s)}.
\label{Eq:sum_eta}
\end{equation}
The ordered motif spectrum is more fine-grained than that of unordered 3-node motifs: dissimilarities in the appearance of ordered motifs within the same (unordered) motif-class are merged and cannot be detected in standard motif analysis (Figure~\ref{figMotifsMap}, top row).
Since the frequencies of different substructures can easily differ by orders of magnitude, we usually plot the motif spectrum on a logarithmic scale.

\section*{A Random Model for Directed Ordered Networks}

We first consider a simple statistical model, the directed ordered random network, that provides a simple approach to generate statistical ensembles of ordered networks and allows to derive properties of the motif spectrum analytically.
Our model is a straightforward generalization of the Erd{\"o}s-R{\'e}nyi random graph model \cite{Newman2010} and depends only on two connection parameters, $p_{\uparrow}$ and $p_{\downarrow}$.
Let $N$ be the number of indexed nodes (we assume that the nodes are ordered according to their rank, $i=1\dots N$).
For each ordered pair of nodes, $j<i$, an upward directed link is introduced with probability $p_{\uparrow}$ from the node $j$ with smaller rank to the node $i$ with larger rank, and a downward directed link is introduced with probability $p_{\downarrow}$.
Thus, for any ordered pair of nodes the probability for the appearance of each of the four possible configurations in Figure~\ref{PossibilitiesTwoOrderedDirected} is given by
\begin{equation}
\begin{array}{ll}
p_1 = (1-p_{\uparrow}) (1-p_{\downarrow}),&\\
p_2 = (1-p_{\uparrow}) p_{\downarrow}, &\\
p_3 = p_{\uparrow} (1-p_\downarrow), &\\
p_4 = p_{\uparrow} p_\downarrow. &
\end{array}
\end{equation}

In a similar way, the  probability for the appearance of each ordered substructure with three nodes (Figure~\ref{figMotifsMap}) can be calculated.
For the substructure $(1,1)$, for example, the appearance probability is
\begin{equation}
P_{(1,1)} = (1-p_{\downarrow}) (1-p_{\uparrow}) p_{\downarrow} (1-p_\uparrow) p_{\downarrow} (1-p_\uparrow) , \\
\label{eqProbability}
\end{equation}
where the first two multipliers describe the absence of links between middle and top nodes, the third and fourth multipliers correspond to the link between bottom and top nodes, and the last two multipliers encode the link between bottom and middle nodes.
The expected number, $\eta_{(q,s)}$, of appearances of each substructure with class ID $q$ and member ID $s$ can be calculated as
\begin{equation}
\eta_{(q,s)} = P_{(q,s)} C_3^N ,
\label{eqNumberAppear}
\end{equation}
where the binomial coefficient $C_3^N = \frac{1}{6} N (N-1) (N-2)$ corresponds to the number of all possible ordered 3-node combinations.

\begin{figure}[ht]
\centering
\includegraphics[width=14cm,keepaspectratio=true,clip=true]{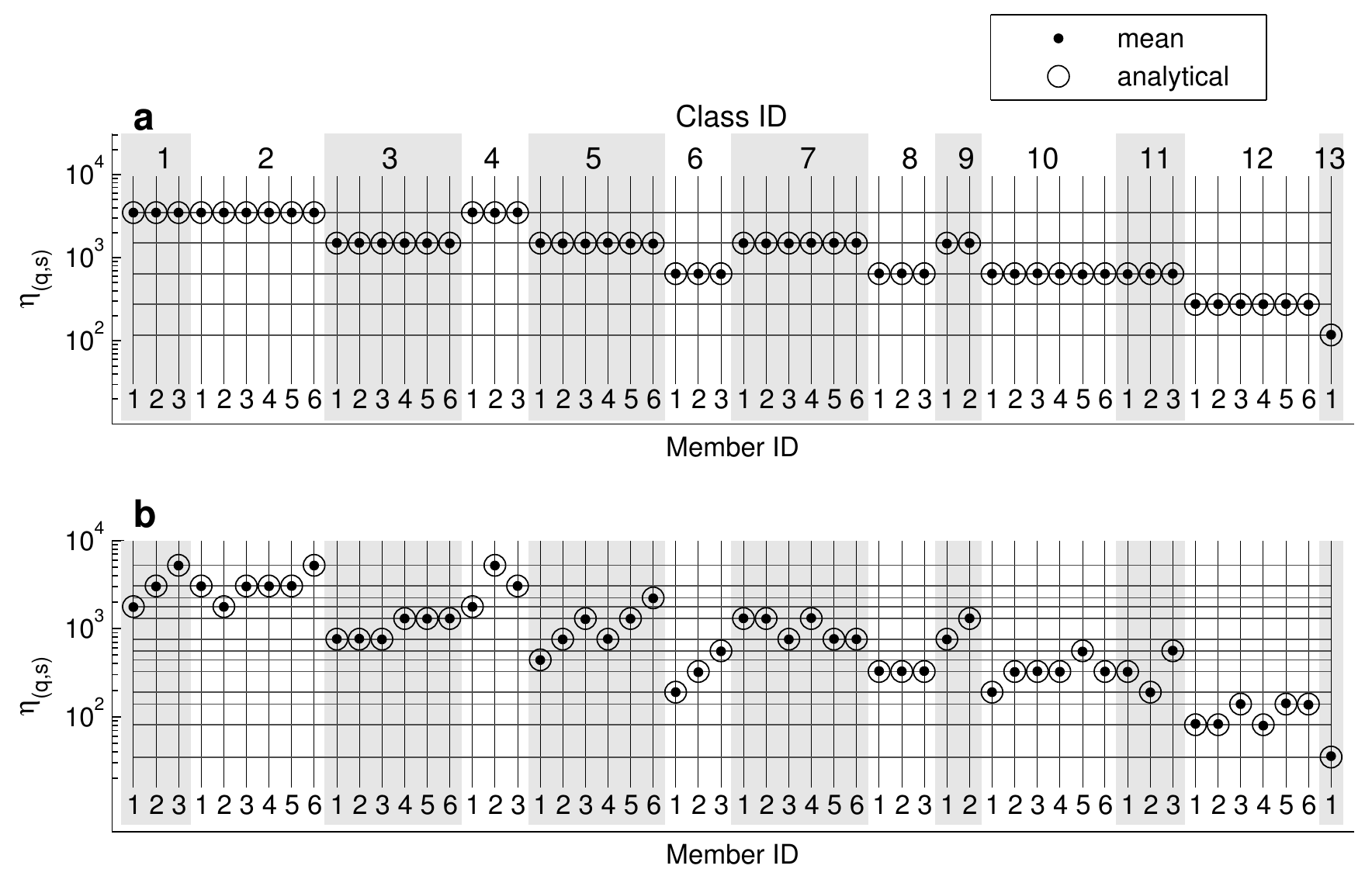}
\caption{
Motif-spectra for the directed ordered random network model.
The plots show the appearance number, $\eta_{(q,s)}$, of ordered 3-node substructures
from numerical simulation of $100$~network realizations with $N=100$  (black dots) and from analytical calculation (open circles).
Note, the logarithmic scale.
The horizontal lines indicate discrete levels of substructure frequencies.
{\bf a}~Symmetric case with $p_\uparrow = p_\downarrow = 0.3$. 
Five distinct levels of appearance numbers according to equation~(\ref{stat5classes}) emerge.   
All substructures within the same motif class have identical appearance numbers.
{\bf b}~Asymmetric case with $p_\uparrow = 0.3$, $p_\downarrow =0.2$. 
Substructures within the same motif class can have different appearance numbers, but are restricted to $13$ distinct levels.
} 
\label{figDiRaGra}
\end{figure}

Figure~\ref{figDiRaGra}a shows an exemplary motif spectrum in the special case of top-down symmetry, where ${p_{\uparrow} = p_{\downarrow} =:p}$. 
In this case the $54$ ordered motifs can be grouped into five sets, each of which has a specific appearance probability (see solid horizontal lines in Figure~\ref{figDiRaGra}a).
Thereby, the probability of appearance for each substructure depends only on the number of directed links~$l$ (with $2 \le l \le 6$) within the motif class\cite{Itzkovitz2005}
\begin{equation}
P_l = p^l (1-p)^{6-l}.
\label{stat5classes}
\end{equation}
Since all substructures of the same motif class have the same number of directed links (Figure~\ref{figMotifsMap}) they also have the same appearance probability.

Next, we study the situation where the top-down symmetry is broken.
Without loss of generality we assume that upward directed links are more common than downward links ($p_{\uparrow} > p_{\downarrow}$).
The corresponding motif spectrum is illustrated in Figure~\ref{figDiRaGra}b. 
In contrast to the symmetric case (Figure~\ref{figDiRaGra}a) the substructures within a single motif class may now have different probabilities of appearance.
Nevertheless, all appearance probabilities in the spectrum are grouped to $13$ distinct levels (marked by solid horizontal lines).
Those levels arise from permutations of multipliers in equation~(\ref{eqProbability}), corresponding to different substructures, but identical probabilities.
We call the set of substructures with identical mean frequencies a {\em statistical class}.
Note that the $13$ statistical classes differ from the $13$ isomorphism classes (Figure~\ref{figMotifsMap}, top row). 
In other words, members of any isomorphism class can differ widely in their rate of appearance.

Remarkably, the spectrum in Figure~\ref{figDiRaGra}b exhibits three dominant substructures,
which correspond to the ordered motifs that contain two upward connections and no downward connection: substructure $(1,3)$, where the node of 
largest rank is reached from upward links by the other two nodes; substructure $(2,6)$, an upward chain; and substructure $(4,2)$, where the node with the smallest rank has an upward link to each of the other nodes.
As will be shown below, the same substructures play an important role in natural food webs.

\section*{Motif analysis of Empirical and Simulated Food webs}

\subsection*{Using the Niche Model to Generate Directed Ordered Networks}

In the following, we show that motif spectra of ordered networks can be used to analyze food web data.
For the analysis we compare data from an empirical lake food web with statistical ensembles of directed ordered networks, which are generated by the niche model\cite{Williams2000}.
The niche model combines stochastic elements with simple link assignment rules and is well known to be able to synthesize networks of trophic interactions between species that closely resemble empirical food webs\cite{Williams2000,Dunne2009}.
The model depends to two parameters, the species richness $N$ (i.e., the number of biological species, each represented by a node in the food web) and the connectance $C$ (i.e., the proportion of possible links in the food web that actually occur). 
In the niche model, each species $i = 1\dots N$ is assigned a random niche variable $n_i \in [0,1]$,  drawn from a uniform distribution.
The niche variable can be regarded to be a proxy of body-size, it determines the outgoing links from this node, and it constitutes a natural ordering of the network nodes as described above.
In the model, species are constrained to consume prey from a contiguous range of species
on the niche axis\cite{Williams2000}.
That is, species $i$ preys upon all species $j$ that have a niche parameter $n_j$ inside a finite segment of length $r_i=x n_i$, centred at a position $c_i$ that is chosen randomly inside the interval $[r_i/2, n_i]$. 
Here, $0 \leq x \leq 1$ is a random variable from a beta distribution $p(x) = \beta (1-x)^{1-\beta}$, with $\beta = (1/2C)-1$.

\subsection*{Motif Spectra of Empirical and Simulated Food Webs}

\begin{figure}[ht]
\centering
\includegraphics[width=14cm,keepaspectratio=true,clip=true]{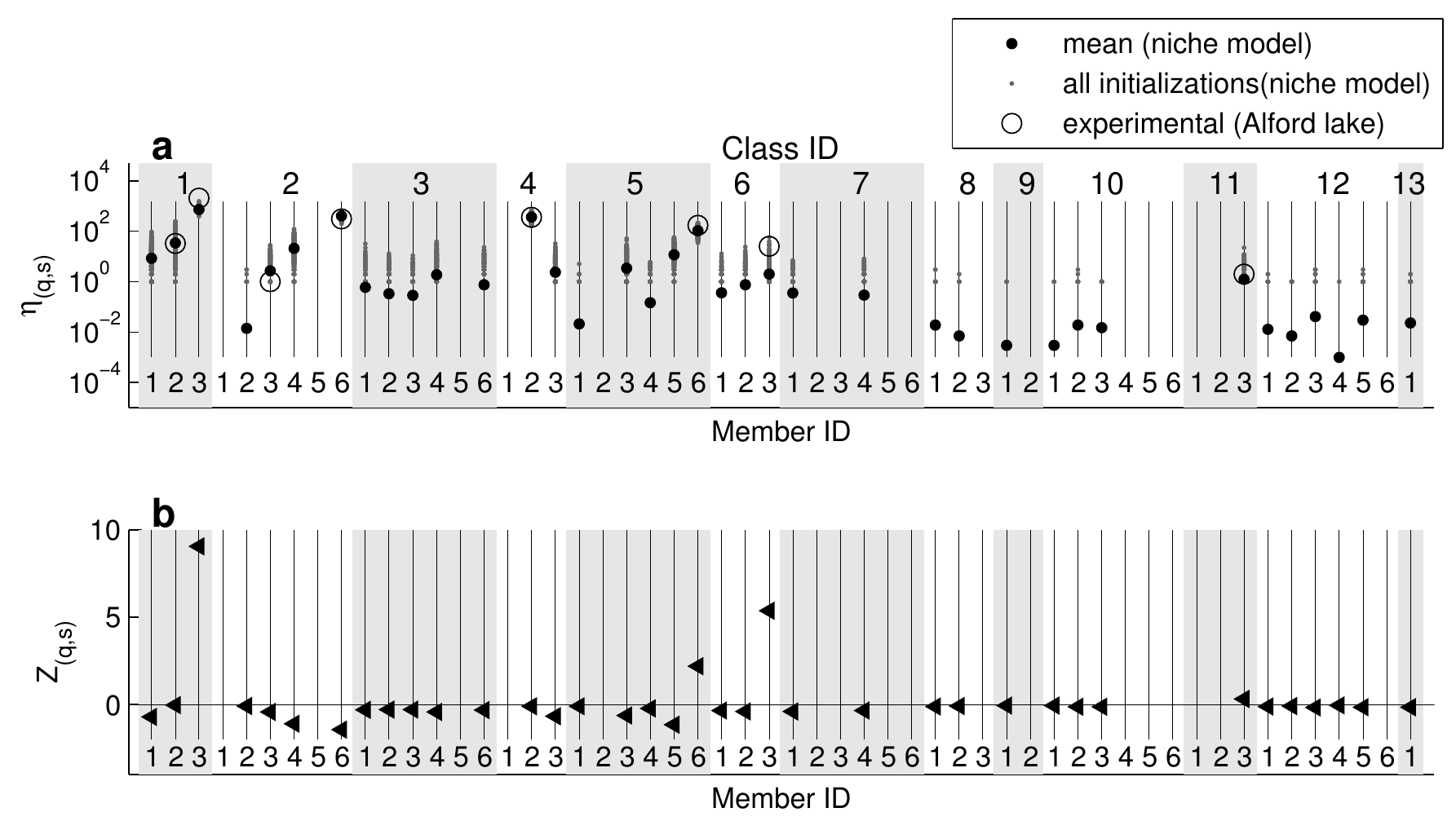}
\caption{
Motif-spectra of empirical and computer generated food-webs ($N = 56$ species, connectance $C = 0.0692$).
{\bf a}~Appearance numbers, $\eta_{(q,s)}$, of ordered 3-node substructures
from the pelagic food web of Alford lake (open circles) and from $1000$ realizations of the fitted niche model (grey dots) on a logarithmic scale.
The mean appearance number of the model realizations is shown by black dots.
Out of the 54 possible substructures, 
in the food web only 8, and in the realizations of the niche model only 37, substructures occur.
{\bf b}~Match between model and experimental data by the Z-score $Z_{(q,s)}$.
Motifs~$(1,3)$, $(6,3)$, and $(5,6)$ show the largest deviation.
}
\label{figHavensNiche}
\end{figure}

In Figure~\ref{figHavensNiche} we compare the motif spectrum of an empirically measured food web with that from an ensemble of computer generated ordered networks. 
Figure~\ref{figHavensNiche}a shows the appearance numbers $\eta_{(q,s)}$ of all ordered 3-node substructures, which we obtained from the pelagic food web of Alford lake from the Adirondack park\cite{Havens1992} (data kindly provided by U.~Brose). 
The average body-mass of adult individuals was used as niche variable to order species.
The data set contains $N = 56$ species with a connectance of $C = 0.0692$.
As shown in the figure, only eight different substructures occur in the empirical motif spectrum.
The most abundant of these are the substructures $(1,3)$, $(2,6)$, and $(4,2)$.
Substructure $(1,3)$ corresponds to a motif of a predator that feeds on two prey species
of smaller body-size 
$(2,6)$ is a tri-trophic chain where prey have smaller body-size than predators, and $(4,2)$ describes a prey that is preyed upon by two predators of higher body-size.
The same substructures have also been identified as the most dominant substructures in the asymmetric random model of directed ordered networks (Figure~\ref{figDiRaGra}b).
This agreement can be explained by the fact that feeding relations between two species are 
not symmetric with respect to body-size.
It is more likely that an upward link from a prey of smaller body-size to a larger predator occurs, than vice versa 
(as usual, we assume that the direction of feeding links in a food web
reflects the direction of energy flow, i.e., a directed link from species A to species B means that B eats A). 

Next, we used the empirical values of $N$ and $C$ to generate statistical food web ensembles with the niche model.
In Figure~\ref{figHavensNiche}a we plot the ordered motif spectra from 1000 realization of the fitted niche model. 
As shown in the figure, in the statistical ensemble we observe a total of $37$ different substructures---clearly more than in the empirical data set.
The substructures that occur in both spectra have appearance frequencies that match rather well and, in general, coincide with the substructures of higher frequencies in the statistical ensemble (Figure~\ref{figHavensNiche}).

To characterize deviations between the model and the empirical data, we use the Z-score
\begin{equation}
Z_{(q,s)} = \frac{\eta_{(q,s)}^{e} - \eta_{(q,s)}^{m}} {\sigma_{(q,s)}^{m}}, 
\label{zscores}
\end{equation}
where $\eta_{(q,s)}^{e}$ denotes the frequency of substructure~$(q,s)$ in the empirical food web, $\eta_{(q,s)}^{m}$ and $\sigma_{(q,s)}^{m}$ are the model mean frequency and standard deviation of a substructure, respectively.
As shown in Figure~\ref{figHavensNiche}b, the strongest deviations between data and model (i.e., a positive Z-score) occur for the motifs $(1,3)$, $(6,3)$, and $(5,6)$,
all of which are connected to a pattern of omnivory (see Fig.~\ref{figMotifsMap}).
These deviations can be explained by the observation that the fitted niche model generates fewer of these motifs than observed in the field. 
The, by far, largest Z-score occurs for the motif $(1,3)$, which is also the most abundant in the empirical motif spectrum (Figure~\ref{figHavensNiche}a).
Analyzing further food-webs of the Adirondack park (results not shown) we find that this is a common pattern: typically the motif spectrum of the niche model deviates most strongly from that of the most abundant substructure in the empirical data.
These results conform with the well-known observation that even though structural food web models,
such as the niche model, are able to provide detailed understanding about the structural complexity of natural food webs, they still show some systematic deficiencies to predict the fine structure of complex food webs\cite{Stouffer2007,Williams2008}.

\begin{figure}[ht]
\centering
\includegraphics[width=14cm,keepaspectratio=true,clip=true]{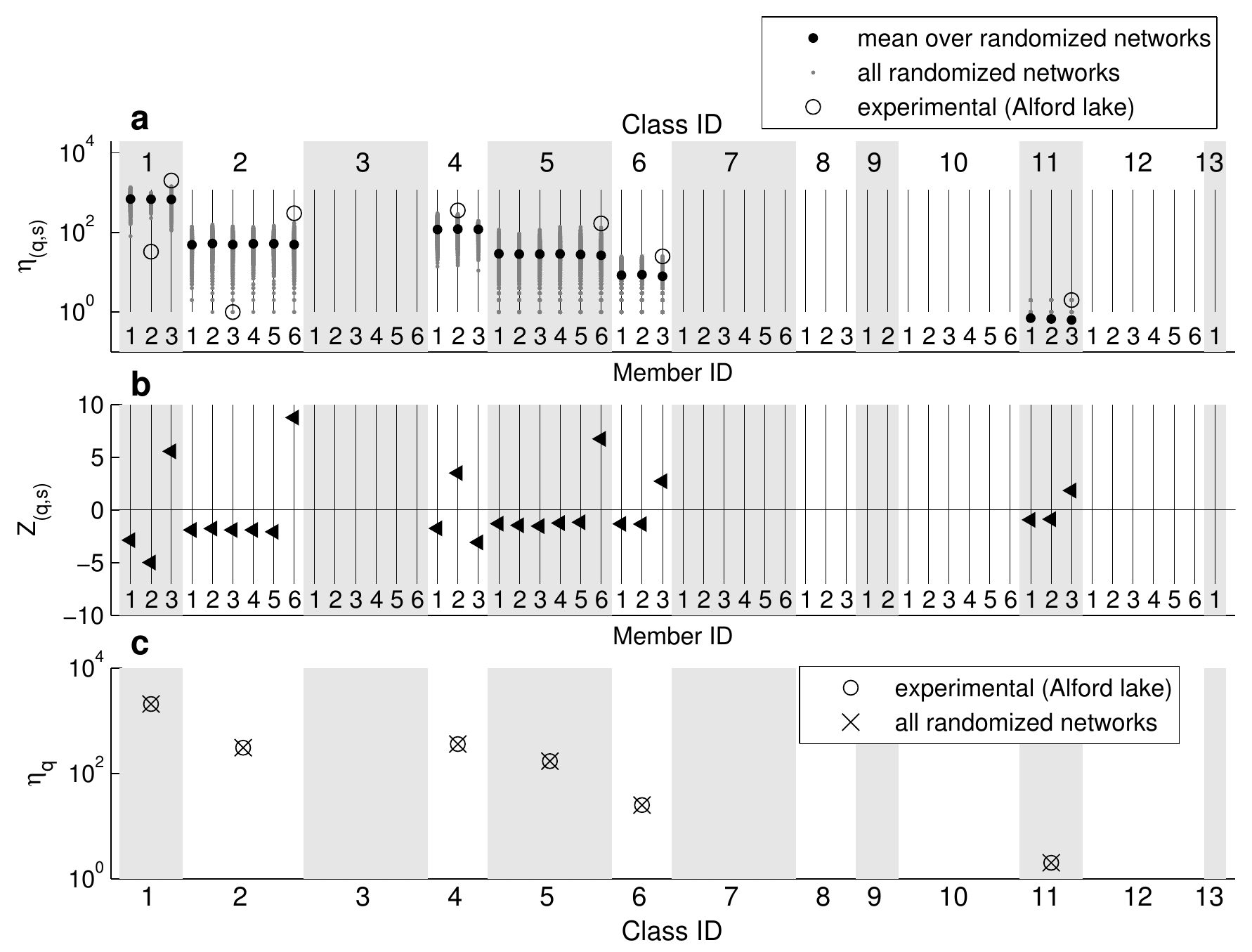}
\caption{
Motif-spectra of the pelagic food-web of Alford lake and its $1000$ randomly reordered realizations (grey dots).     
Data can only be shown for the 6 motif classes that actually occur in the empirical food web.
{\bf a}~Comparison of the absolute motif counts $\eta_{(q,s)}$  on a logarithmic scale.
{\bf b}~Relative deviations between the empirical and randomized food webs, measured by the $Z$-score $Z_{(q,s)}$.
{\bf c}~Comparison of the number $\eta_q$ of unordered substructures 
according to equation (\ref{Eq:sum_eta}) between experimental data and randomized networks on a logarithmic scale. 
Symbols are explained in the legend. }
\label{figHavens1Permutation}
\end{figure}

To test for the relevance of body-size ordering in empirical food webs, we compare the motif spectrum of the pelagic food web of Alford lake with that of its randomly re-ordered counterparts (Figure~\ref{figHavens1Permutation}a). 
If a substructure does not appear in the empirical food web we set $\eta_{(q,s)} = 0$ (not shown on the logarithmic scale in Figure~\ref{figHavens1Permutation}a).
The figure reveals only slight agreement between the motif spectra of the empirical food webs before and after randomized ordering.
This is confirmed by our calculation of the spectrum of Z-scores in Figure~\ref{figHavens1Permutation}b.
Note, that now $\eta_{(q,s)}^{m}$ and $\sigma_{(q,s)}^{m}$ in Eq.~(\ref{zscores}) denote the mean frequency and standard deviation of a substructure in the randomized webs.
The large entries in the spectrum of Z-scores reveal substantial deviations in the structure of the natural and randomized food webs.
If the body-size order is neglected by summing over all member IDs, the spectrum reduces to the $13$ standard unordered motif classes and the spectra of empirical and randomized networks become indistinguishable (Figure~\ref{figHavens1Permutation}c).
These results indicate that hierarchy due to body-size is a crucial aspect of the structure of empirical food webs.
The precise role of body-size order for structuring natural food webs
provides an intriguing possibility for future research.

\section*{Discussion}

The method presented in this paper is a natural extension of the classic network motif analysis\cite{Milo2002} to networks with hierarchically structured nodes.
For these networks, the spectrum of ordered substructures yields a quantitative description of the connectivity-patterns with respect to node-rank.
Thereby, highly abundant ordered motifs, such as substructure $(1,3)$ in Figure~\ref{figDiRaGra}b or $(4,2)$ in Figure~\ref{figHavensNiche}, represent connectivity-patterns that are typical within the node hierarchy.
The spectrum can be expressed in absolute motif counts~$\eta_{(q,s)}$ or appearance probabilities~$P_{(q,s)}$, whereby the latter is independent of network size, which makes it suitable for unified comparisons across different networks.

We have shown that ordered motif spectra can reflect a breaking of the top-down symmetry
in the hierarchy among nodes.
This means that networks, in which connections to nodes of higher rank are more frequent than downward directed connections, naturally contain a larger share of the corresponding motifs that are mostly composed of upward links, or vice versa.
In general, we observe large variations in the frequencies of ordered substructures that easily can span several orders of magnitudes.
A high abundance of an ordered motif does, however, not directly imply its statistical significance.
The latter requires a model to compare against, as demonstrated by our use of the randomly reordered networks and the niche model.
On the other hand, rare motifs should not immediately be disregarded, as they might still play a key role in the network.
Assume, for example, that in a food web most of the biomass is concentrated on a small number of interacting species.
In this case, even if the corresponding motif(s) are very rare, they could still carry an important ecosystem function.
Therefore, in general, we suggest to take the whole spectrum of ordered motifs into account for network analysis.

In this work we only considered substructures composed of three nodes, 
however the presented technique generalizes to larger motifs.
In practice, counting larger motifs might be more challenging due to the rapidly increasing computational demands with growing motif- and network-size.
Our approach also generalizes to different classes of natural and theoretical networks.
Here, we have discussed food webs, using the species' niche coordinates to define the ordering among the network nodes.
But many other networks, for which motifs have been analyzed traditionally\cite{Milo2002}, might possess natural ordering criteria or hidden niche variables, which would allow for a worthwhile reanalysis using ordered motifs.

Finally, the proposed approach may be helpful to detect structural changes in adaptive networks that are coevolutionary changing in time (e.g., evolving food webs)\cite{Gross2008}
and it might open new avenues for generalizing motif analysis to 
networks that contain multiple layers of connectivity\cite{Gao2012,Kivela2014}.
Such structures recently have gotten in focus of the scientific attention as networks of networks or multi-layered networks and are characterized by nodes that are connected by more than one type of relationship.
For example, the risk of cascading failures in important infrastructure facilities, such as  electrical power grids \cite{Albert2004}, may be related to connections within multiple interdependent communication, transport, or infrastructure subsystems \cite{Gao2012}.
Similar, in ecology trophic interactions represent only one of many 
possible forms by which species can influence each other.
It is increasingly recognized that ecological networks
contain different interactions beyond feeding relationships, such as 
host-parasitoid interactions, interference competition, 
and other forms of non-trophic interactions (e.g., mutualism, habitat modification, or facilitation)
\cite{Ings2009,Kefi2015}.
The exploration of the structural and dynamical properties of such multi-layered networks is still in its infancy
and in particular, simple robust techniques are needed 
that allow to capture the huge complexity of such systems\cite{Kivela2014}. 
In this sense, our proposed method may not only be helpful to identify hierarchical patterns in empirical food webs and other natural networks, it may also provide the base for extending motif analysis to multi-layered networks.

%


\begin{thebibliography}{10}
\expandafter\ifx\csname url\endcsname\relax
  \def\url#1{\texttt{#1}}\fi
\expandafter\ifx\csname urlprefix\endcsname\relax\def\urlprefix{URL }\fi
\providecommand{\bibinfo}[2]{#2}
\providecommand{\eprint}[2][]{\url{#2}}

\bibitem{Newman2010}
\bibinfo{author}{Newman, M.}
\newblock \emph{\bibinfo{title}{Networks: an introduction}}
  (\bibinfo{publisher}{Oxford University Press}, \bibinfo{year}{2010}).

\bibitem{Blasius2011}
\bibinfo{author}{Blasius, B.} \& \bibinfo{author}{Brockmann, D.}
\newblock \bibinfo{title}{Frontiers in network science: advances and
  applications}.
\newblock \emph{\bibinfo{journal}{EPJ B}} \textbf{\bibinfo{volume}{84}},
  \bibinfo{pages}{491--492} (\bibinfo{year}{2011}).

\bibitem{Borgatti2009}
\bibinfo{author}{Borgatti, S.~P.}, \bibinfo{author}{Mehra, A.},
  \bibinfo{author}{Brass, D.~J.} \& \bibinfo{author}{Labianca, G.}
\newblock \bibinfo{title}{Network analysis in the social sciences}.
\newblock \emph{\bibinfo{journal}{Science}} \textbf{\bibinfo{volume}{323}},
  \bibinfo{pages}{892--895} (\bibinfo{year}{2009}).

\bibitem{Jackson2008}
\bibinfo{author}{Jackson, M.~O.} \emph{et~al.}
\newblock \emph{\bibinfo{title}{Social and economic networks}},
  vol.~\bibinfo{volume}{3} (\bibinfo{publisher}{Princeton University Press
  Princeton}, \bibinfo{year}{2008}).

\bibitem{Albert2004}
\bibinfo{author}{Albert, R.}, \bibinfo{author}{Albert, I.} \&
  \bibinfo{author}{Nakarado, G.~L.}
\newblock \bibinfo{title}{Structural vulnerability of the north american power
  grid}.
\newblock \emph{\bibinfo{journal}{Phys. Rev. E}}
  \textbf{\bibinfo{volume}{69}}, \bibinfo{pages}{025103}
  (\bibinfo{year}{2004}).

\bibitem{Kaluza2010}
\bibinfo{author}{Kaluza, P.}, \bibinfo{author}{K{\"o}lzsch, A.},
  \bibinfo{author}{Gastner, M.~T.} \& \bibinfo{author}{Blasius, B.}
\newblock \bibinfo{title}{The complex network of global cargo ship movements}.
\newblock \emph{\bibinfo{journal}{J. R. Soc. Interface}} \textbf{\bibinfo{volume}{7}}
  \bibinfo{pages}{1093--1103} (\bibinfo{year}{2010}).

\bibitem{Alon2003}
\bibinfo{author}{Alon, U.}
\newblock \bibinfo{title}{Biological networks: the tinkerer as an engineer}.
\newblock \emph{\bibinfo{journal}{Science}} \textbf{\bibinfo{volume}{301}},
  \bibinfo{pages}{1866--1867} (\bibinfo{year}{2003}).

\bibitem{Jordan2004}
\bibinfo{author}{Jord{\'a}n, F.} \& \bibinfo{author}{Scheuring, I.}
\newblock \bibinfo{title}{Network ecology: topological constraints on ecosystem
  dynamics}.
\newblock \emph{\bibinfo{journal}{Phys. Life Rev.}}
  \textbf{\bibinfo{volume}{1}}, \bibinfo{pages}{139--172}
  (\bibinfo{year}{2004}).

\bibitem{Proulx2005}
\bibinfo{author}{Proulx, S.~R.}, \bibinfo{author}{Promislow, D.~E.} \&
  \bibinfo{author}{Phillips, P.~C.}
\newblock \bibinfo{title}{Network thinking in ecology and evolution}.
\newblock \emph{\bibinfo{journal}{Trends Ecol. Evol.}}
  \textbf{\bibinfo{volume}{20}}, \bibinfo{pages}{345--353}
  (\bibinfo{year}{2005}).

\bibitem{Dunne2009}
\bibinfo{author}{Dunne, J.~A.}
\newblock \bibinfo{title}{Food webs}.
\newblock In \bibinfo{editor}{Meyers, R.~A.} (ed.)
  \emph{\bibinfo{booktitle}{Encyclopedia of Complexity and Systems Science}},
  \bibinfo{pages}{3661--3682} (\bibinfo{publisher}{Springer},
  \bibinfo{year}{2009}).

\bibitem{Costa2007}
\bibinfo{author}{Costa, L. d.~F.}, \bibinfo{author}{Rodrigues, F.~A.},
  \bibinfo{author}{Travieso, G.} \& \bibinfo{author}{Villas~Boas, P.~R.}
\newblock \bibinfo{title}{Characterization of complex networks: A survey of
  measurements}.
\newblock \emph{\bibinfo{journal}{Adv. Phys.}}
  \textbf{\bibinfo{volume}{56}}, \bibinfo{pages}{167--242}
  (\bibinfo{year}{2007}).

\bibitem{Milo2002}
\bibinfo{author}{Milo, R.} \emph{et~al.}
\newblock \bibinfo{title}{Network motifs: Simple building blocks of complex
  networks}.
\newblock \emph{\bibinfo{journal}{Science}} \textbf{\bibinfo{volume}{298}},
  \bibinfo{pages}{824--827} (\bibinfo{year}{2002}).

\bibitem{Milo2004}
\bibinfo{author}{Milo, R.} \emph{et~al.}
\newblock \bibinfo{title}{Superfamilies of evolved and designed networks}.
\newblock \emph{\bibinfo{journal}{Science}} \textbf{\bibinfo{volume}{303}},
  \bibinfo{pages}{1538--1542} (\bibinfo{year}{2004}).

\bibitem{Sporns2004}
\bibinfo{author}{Sporns, O.} \& \bibinfo{author}{K\"otter, R.}
\newblock \bibinfo{title}{Motifs in brain networks}.
\newblock \emph{\bibinfo{journal}{PLoS Biol.}} \textbf{\bibinfo{volume}{2}},
  \bibinfo{pages}{e369} (\bibinfo{year}{2004}).

\bibitem{Itzkovitz2005}
\bibinfo{author}{Itzkovitz, S.} \& \bibinfo{author}{Alon, U.}
\newblock \bibinfo{title}{Subgraphs and network motifs in geometric networks}.
\newblock \emph{\bibinfo{journal}{Phys. Rev. E}} \textbf{\bibinfo{volume}{71}},
  \bibinfo{pages}{026117} (\bibinfo{year}{2005}).

\bibitem{Stouffer2007}
\bibinfo{author}{Stouffer, D.~B.}, \bibinfo{author}{Camacho, J.},
  \bibinfo{author}{Jiang, W.} \& \bibinfo{author}{Amaral, L. A.~N.}
\newblock \bibinfo{title}{Evidence for the existence of a robust pattern of
  prey selection in food webs}.
\newblock \emph{\bibinfo{journal}{Proc. R. Soc. B}}
  \textbf{\bibinfo{volume}{274}}, \bibinfo{pages}{1931--1940} (\bibinfo{year}{2007}).

\bibitem{Caldarelli2002}
\bibinfo{author}{Caldarelli, G.}, \bibinfo{author}{Capocci, A.},
  \bibinfo{author}{De~Los~Rios, P.} \& \bibinfo{author}{Mu\~noz, M.~A.}
\newblock \bibinfo{title}{Scale-free networks from varying vertex intrinsic
  fitness}.
\newblock \emph{\bibinfo{journal}{Phys. Rev. Lett.}}
  \textbf{\bibinfo{volume}{89}}, \bibinfo{pages}{258702}
  (\bibinfo{year}{2002}).

\bibitem{Williams2000}
\bibinfo{author}{Williams, R.~J.} \& \bibinfo{author}{Martinez, N.~D.}
\newblock \bibinfo{title}{Simple rules yield complex food webs}.
\newblock \emph{\bibinfo{journal}{Nature}} \textbf{\bibinfo{volume}{404}},
  \bibinfo{pages}{180--183} (\bibinfo{year}{2002}).

\bibitem{Williams2008}
\bibinfo{author}{Williams, R.~J.} \& \bibinfo{author}{Martinez, N.~D.}
\newblock \bibinfo{title}{Success and its limits among structural models of
  complex food webs}.
\newblock \emph{\bibinfo{journal}{J. Anim. Ecol.}}
  \textbf{\bibinfo{volume}{77}}, \bibinfo{pages}{512--519} (\bibinfo{year}{2008}).

\bibitem{Williams2011}
\bibinfo{author}{Williams, R.~J.} \& \bibinfo{author}{Purves, D.~W.}
\newblock \bibinfo{title}{The probabilistic niche model reveals substantial
  variation in the niche structure of empirical food webs}.
\newblock \emph{\bibinfo{journal}{Ecology}} \textbf{\bibinfo{volume}{92}},
  \bibinfo{pages}{1849--1857} (\bibinfo{year}{2011}).

\bibitem{Havens1992}
\bibinfo{author}{Havens, K.}
\newblock \bibinfo{title}{Scale and structure in natural food webs}.
\newblock \emph{\bibinfo{journal}{Science}} \textbf{\bibinfo{volume}{257}},
  \bibinfo{pages}{1107--1109} (\bibinfo{year}{1992}).

\bibitem{Condit2000}
\bibinfo{author}{Condit, R.} \emph{et~al.}
\newblock \bibinfo{title}{Spatial patterns in the distribution of tropical tree
  species}.
\newblock \emph{\bibinfo{journal}{Science}} \textbf{\bibinfo{volume}{288}},
  \bibinfo{pages}{1414--1418} (\bibinfo{year}{2000}).

\bibitem{Tuomisto2003}
\bibinfo{author}{Tuomisto, H.}, \bibinfo{author}{Ruokolainen, K.} \&
  \bibinfo{author}{Yli-Halla, M.}
\newblock \bibinfo{title}{Dispersal, environment, and floristic variation of
  western amazonian forests}.
\newblock \emph{\bibinfo{journal}{Science}} \textbf{\bibinfo{volume}{299}},
  \bibinfo{pages}{241--244} (\bibinfo{year}{2003}).

\bibitem{Thieltges2009}
\bibinfo{author}{Thieltges, D.~W.} \emph{et~al.}
\newblock \bibinfo{title}{Distance decay of similarity among parasite
  communities of three marine invertebrate hosts}.
\newblock \emph{\bibinfo{journal}{Oecologia}} \textbf{\bibinfo{volume}{160}},
  \bibinfo{pages}{163--173} (\bibinfo{year}{2009}).

\bibitem{Lang2013}
\bibinfo{author}{Lang, B.}, \bibinfo{author}{Rall, B.~C.},
  \bibinfo{author}{Scheu, S.} \& \bibinfo{author}{Brose, U.}
\newblock \bibinfo{title}{Effects of environmental warming and drought on
  size-structured soilfood webs}.
\newblock \emph{\bibinfo{journal}{OIKOS}} \textbf{\bibinfo{volume}{123}},
  \bibinfo{pages}{1224--1233} (\bibinfo{year}{2014}).

\bibitem{Loeuille2005}
\bibinfo{author}{Loeuille, N.} \& \bibinfo{author}{Loreau, M.}
\newblock \bibinfo{title}{Evolutionary emergence of size-structured food webs}.
\newblock \emph{\bibinfo{journal}{PNAS}} \textbf{\bibinfo{volume}{102}},
  \bibinfo{pages}{5761--5766} (\bibinfo{year}{2005}).

\bibitem{Allhoff2013}
\bibinfo{author}{Allhoff, K.~T.} \& \bibinfo{author}{Drossel, B.}
\newblock \bibinfo{title}{When do evolutionary food web models generate complex
  networks?}
\newblock \emph{\bibinfo{journal}{J. Theor. Biol.}}
  \textbf{\bibinfo{volume}{334}}, \bibinfo{pages}{122--129} (\bibinfo{year}{2013}).

\bibitem{CamachoStoufferAmaral2007}
\bibinfo{author}{Camacho, J.}, \bibinfo{author}{Stouffer, D.} \&
  \bibinfo{author}{Amaral, L. A.~N.}
\newblock \bibinfo{title}{Quantitative analysis of the local structure of food
  webs}.
\newblock \emph{\bibinfo{journal}{J. Theor. Biol.}}
  \textbf{\bibinfo{volume}{246}}, \bibinfo{pages}{260--268} (\bibinfo{year}{2007}).

\bibitem{Stouffer2010}
\bibinfo{author}{Stouffer, D.~B.}
\newblock \bibinfo{title}{Scaling from individuals to networks in food webs}.
\newblock \emph{\bibinfo{journal}{Funct. Ecol.}} \textbf{\bibinfo{volume}{24}},
  \bibinfo{pages}{44--51} (\bibinfo{year}{2010}).

\bibitem{Gross2008}
\bibinfo{author}{Gross, T.} \& \bibinfo{author}{Blasius, B.}
\newblock \bibinfo{title}{Adaptive coevolutionary networks: a review}.
\newblock \emph{\bibinfo{journal}{J. R. Soc. Interface}}
  \textbf{\bibinfo{volume}{5}}, \bibinfo{pages}{259--271}
  (\bibinfo{year}{2008}).

\bibitem{Gao2012}
\bibinfo{author}{Gao, J.}, \bibinfo{author}{Buldyrev, S.~V.},
  \bibinfo{author}{Stanley, H.~E.} \& \bibinfo{author}{Havlin, S.}
\newblock \bibinfo{title}{Networks formed from interdependent networks}.
\newblock \emph{\bibinfo{journal}{Nature Phys.}}
  \textbf{\bibinfo{volume}{8}}, \bibinfo{pages}{40--48} (\bibinfo{year}{2012}).

\bibitem{Kivela2014}
\bibinfo{author}{Kivel{\"a}, M.} \emph{et~al.}
\newblock \bibinfo{title}{Multilayer networks}.
\newblock \emph{\bibinfo{journal}{J. Complex Networks}}
  \textbf{\bibinfo{volume}{2}}, \bibinfo{pages}{203--271}
  (\bibinfo{year}{2014}).

\bibitem{Ings2009}
\bibinfo{author}{Ings, T.~C.} \emph{et~al.}
\newblock \bibinfo{title}{Review: Ecological networks--beyond food webs}.
\newblock \emph{\bibinfo{journal}{J. Anim. Ecol.}}
  \textbf{\bibinfo{volume}{78}}, \bibinfo{pages}{253--269}
  (\bibinfo{year}{2009}).

\bibitem{Kefi2015}
\bibinfo{author}{K{\'e}fi, S.} \emph{et~al.}
\newblock \bibinfo{title}{Network structure beyond food webs: mapping
  non-trophic and trophic interactions on chilean rocky shores}.
\newblock \emph{\bibinfo{journal}{Ecology}} \textbf{\bibinfo{volume}{96}},
  \bibinfo{pages}{291--303} (\bibinfo{year}{2015}).

\end{thebibliography}

\section*{Acknowledgements}

We thank D.B.~Stouffer and C.~Guill for inspiring discussions and U.~Brose for providing the experimental data.
We are grateful to German VW-foundation and DFG project FOR~1748 for funding.

\section*{Author contributions statement}

P.V.P. and C.F. conceived the experiments.
P.V.P. performed the experiments.
B.B. supervised the study.
All authors analyzed the results, wrote and reviewed the manuscript.

\section*{Additional information}

PACS numbers: 42.65.Tg; 42.81.Dp; 
The authors declare no competing financial interests.

\newpage

\end{document}